\documentclass[epj]{svjour}
\usepackage{graphicx}
\usepackage{amsmath,amssymb}
\begin{document}
\title{Rheological aging and rejuvenation in solid friction contacts}
\author{L. Bureau, T. Baumberger \and C. Caroli}

\institute{Groupe de Physique des Solides (UMR 7588),
Universit\'es Paris 6 \&
7, 2 place Jussieu, 75251 Paris Cedex 05, France.}
\date{}
\abstract{
We study the low-velocity (0.1--100 $\mu$m.s$^{-1}$) frictional 
properties of interfaces between a rough glassy polymers and smooth 
silanized glass, a configuration which gives direct access to the 
rheology of the adhesive joints in which shear localizes. We show that 
these joints exhibit the full phenomenology expected for confined 
quasi 2D soft glasses: they strengthen logarithmically when aging at 
rest, and weaken (rejuvenate) when sliding. Rejuvenation is found to 
saturate at large velocities. Moreover, aging at rest is shown to be 
strongly accelerated when waiting under finite stress below the static 
threshold.
\PACS{ {}{}
} % end of PACS codes      
} %end of abstract
\maketitle
\section{Introduction}
\label{intro}
Surfaces of macroscopic solids in general present some degree of
roughness on the small --- typically micrometric --- scale. For this
reason, most extended interfaces between two nominally flat solids
are of the multicontact type, i.e. composed of a sparse set of real
contacts between load-bearing asperities (Fig.\ref{fig:1}).
Their typical size lies in
the 1--10 $\mu$m range,  and they bear average normal stresses $\bar
p$ of the order of the yield stress $Y$ of the weaker bulk material 
\cite{BT}.

In frictional sliding, it is in the nanometer-thick
region --- the ``joint'' ---
where the two surfaces come into molecular contact and feel the
short-range adhesive interactions, that shear
localizes, which indicates that the joint is mechanically weaker than
the cohesive bulk \cite{PRB2}.

Following Tabor \cite{BT}, the solid friction force $F$ needed to slide one
solid against the other at relative velocity $\dot x$ reads:
\begin{equation}
     \label{eq:tabor}
     F = \sigma_{s}\Sigma_{r}
\end{equation}
with $\Sigma_{r}$ the total real area of contact, and $\sigma_{s}$
the shear stress within microcontacts.

It is now well established that, at low velocities, the variations of
the friction force are governed by the competition between two
effects:

(i) {\it Geometric} aging: since $\bar p \sim Y$, the load-bearing
asperities creep plastically, hence $\Sigma_{r}$ increases
quasi-log\-arithmically with contact duration $\Phi$, which may thus be
termed ``geometric age''. When the interface is at rest, $\Phi = t$.
When motion starts, contacts
get gradually destroyed, after a lifetime or age $\Phi$, and
replaced by fresh ones. So, while the interface sits still, it
ages (strengthens), when it slides, it rejuvenates (weakens).
This is reflected in the so-called static friction peak observed when
setting the system into motion by loading it at constant velocity 
\cite{PRB1}.

(ii) Velocity dependence of the sliding stress $\sigma_{s}$: For
various types of such rough/rough multicontact interfaces (polymer
glasses \cite{PRB2}, rocks \cite{nakatani,scholz}, paper \cite{PRE94})
the analysis of the sliding dynamics has led
to conclude that, for $\dot x$ typically in the 1--100 $\mu$m.s$^{-1}$
range, and up to experimental accuracy:

\begin{equation}
     \label{eq:rheo}
     \sigma_{s}(\dot x)\cong \sigma_{s0}\left(1+\alpha\ln\frac{\dot
     x}{V_{0}}\right)
\end{equation}
with $V_{0}$ some reference velocity in this range.
For most systems $\alpha$ lies in the $10^{-2}$ range.
This rheology can be interpreted as resulting from thermally
activated ``depinning'' dissipative events occurring within the
adhesive joint and involving activation volumes in the (nm)$^{3}$
range \cite{PRB2,persson}.

This is consistent with the following qualitative description of
adhesive solid friction joints:

(i) A joint is a molecularly disordered, quasi-$2D$ med\-ium with
thickness $h$ of nanometric order. This is obviously the case when the
bulk solids themselves are  amorphous.  It should also be so for a
vast majority of solids ordered in the bulk, due to nanometric
surface roughnesses, defects in the bulk, atomic misorientation
between surfaces, and the frequent presence of adsorbates 
\cite{robbins}.

(ii) At rest under zero shear stress, this highly confined medium is
a solid. Indeed it is found to respond elastically to a small
amplitude shear force.

(iii) When this force reaches the static friction threshold,
$F_{s} = \mu_{s}W$ (with $W$ the normal load and $\mu_{s}$ the static
friction coefficient), sliding sets in.

Such a phenomenology entails that frictional sliding can be understood
as plastic flow of this $2D$ confined medium sheared at a deformation
rate $\dot x/h$ proportional to the interfacial sliding velocity. The
static threshold then corresponds to what is usually called, in the
language of bulk plasticity, the yield stress.

Recent theoretical \cite{sollich1,sollich2,kurchan,lequeux1} and experimental 
\cite{bonn,cloitre,derec} developments in the
% related fields of plasticity of amorphous media, and
rheology of complex fluids
and (e.g. soft) glasses, have brought to light the important fact
that the main characteristics of such rheologies can be interpreted as
resulting from the interplay between:

(i) Glass-like aging: As a quenched amorphous structure relaxes down
its glassy energy landscape, the system becomes mechanically stronger.
This age-strengthening is usually roughly logarithmic.

(ii) Rejuvenation by motion: In the plastic regime, dissipative
events, instead of being solely due to activation by (thermal or dynamical)
noise, are accelerated by the stress-induced biasing of the energy
landscape. The ``young'', shallow energy states which were depopulated
as aging pushed the system into ``older'', deeper ones, thus get
repopulated: The sheared soft glass rejuvenates.

If the characterization of solid friction proposed above is truly
relevant, it should therefore be possible to evidence {\it structural}
aging {\it vs} rejuvenation effects in the rheology of adhesive
joints. If such effects are present, a more direct --- hence
accurate --- experimental study of the fine variations of the
frictional stress should reveal that $\sigma_{s}$ not only depends on
the instantaneous deformation rate $\dot x /h$, but also on the previous
history of the joint.

In this article  we show, on the basis of a detailed experimental
study of the frictional dynamics of a \{rough PMMA/float glass\}
interface, that such is indeed the case. The use of such a
configuration, in which the microcontacts permanently retain their
identity, enables us to get rid of the {\it geometric} aging/rejuvenation
effects which are predominant for rough/rough interfaces. We are thus
able to show that, while an adhesive joint strengthens quasi-logarithmically
as it ages at
rest, it rejuvenates when sliding. Moreover, we report experiments in
which the joint ages  under a finite shear stress below the static
threshold. They clearly demonstrate that, the larger this applied
stress, the faster the aging dynamics of the joint. All this
phenomenology proves that a frictional joint can be considered as a
confined {\it 2D} soft glass.

In section \ref{sec:1} we describe the system and measurement setup,
and the experimental procedures. Section \ref{sec:2} reports the
experimental results, which are discussed in section \ref{sec:3}.

\section{Experimental setup and system preparation}
\label{sec:1}
\subsection{Sample preparation}
\label{subsec:1.1}
The system consists of a slider and a track. The slider is a
poly(methyl methacrylate) [PMMA] block, with thickness $10$ mm and a
nominal area $20\times70$ mm$^{2}$, sliding along its longer dimension. The
sliding surface is roughened by hand-lapping against a flat reference
glass plate with SiC powder (GRIT $400$). The resulting r.m.s.
roughness is $1.3 \, \mu$m. After lapping, the surface is thoroughly
rinsed with distilled water an dried by an Ar flux.

The track is a float soda lime glass $120\times120$ mm$^{2}$ plate, of
thickness $10$ mm, with subnanometric roughness on the micrometer scale.
Its surfaces are prepared as follows :

(i) {\it Cleaning}

The plate, degreased with methanol, ethanol and acetone, is sonicated
first in a $2 \%$ solution of RBS detergent for $15$ minutes
at $50^\circ$C, then in ultrapure milli-Q water. After  drying it by a
N$_{2}$ flux, ultimate cleaning is performed in a UV-ozone reactor
for $2$ hours under the steady O$_{2}$ flux $0.5$ l/min, both surfaces
being successively exposed to the UV light.

We thus obtain what we will call ``bare" glass. In the cases where we
use it as such, the plate is immediately transferred from the clean
room into the glove box in which all experiments are performed.

(ii) {\it Silanization}

Most track samples are silanized before use, with the short molecule
(1,1,1,3,3,3) hexamethyl disilazane, in a vacuum chamber. Prior to
silanization, the plate is exposed for $12$ hours to a saturated water
vapor in order to maximize the number of reactive Si-OH surface sites.
After pumping out this vapor, the evacuated chamber and the flask
containing 1 ml of silane are separately heated up to $90^\circ$C, then put
in communication, which results in total silane evaporation. The
reaction is carried on at $90^\circ$C for either 1 (D1 tracks) or 5
days (D5 tracks).

After cooling, the samples are transferred to the glove box, where we
maintain a small Ar overpressure, in the presence of a desiccant
resulting in 8--10\%  relative humidity. All experiments are
performed at temperature $20^\circ$C, well below the PMMA glass transition
temperature ($T_{g}\simeq 110^{\circ}$C.

\subsection{ Measurement setup}
\label{subsec:1.2}

\begin{figure}[htbp]
    $$
    \includegraphics[width=70mm]{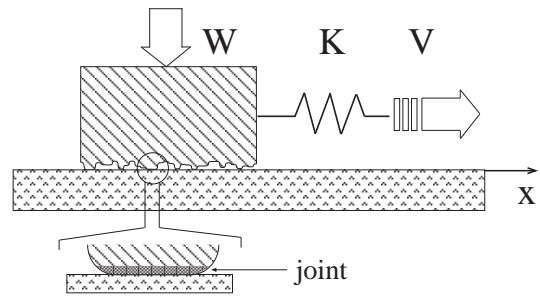}
    $$
    \caption{Experimental setup: the PMMA sliding block, of weight 
    $W$, is loaded through a spring of stiffness $K$ driven at 
    velocity $V$, along the glass track. Blow-up: microcontact 
    between a PMMA asperity and the track.}
    \label{fig:1}
\end{figure}

The mechanical setup has been described in detail elsewhere \cite{PRB1}. It is
schematized on Figure \ref{fig:1}. The normal load on the interface
is provided
by the slider dead weight, $W = 7.7$ N. The slider is driven by a
stepping motor through a loading spring of stiffness
$K = 8 \times 10^{-2}$ N/$\mu$m,
the elongation of which, measured with the help of a
capacitive sensor, yields the tangential force to within $10^{-4}$ N.
Driving velocities range between $10^{-1}$ and $10^{2} \, \mu$m/s.

\subsection{Preparing an interface of constant $\Sigma_{r}$}
\label{subsec:1.3}

\begin{figure}[htbp]
    $$
    \includegraphics[width=70mm]{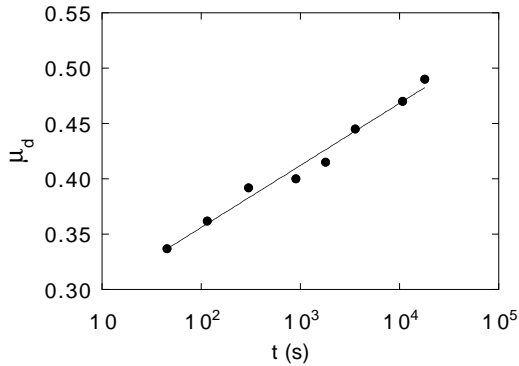}
    $$
    \caption{Evolution, on a bare glass track, of the dynamic friction coefficient $\mu_{d}$ 
    at velocity $V_{0}=50\, \mu$m.s$^{-1}$ with time $t$ elapsed from 
    the creation of the interface. Line: best logarithmic fit.}
    \label{fig:2}
\end{figure}

In a preliminary series of experiments, the slider and track are
brought into contact at an initial instant $t =0$. We then measure the
dynamic friction coefficient $\mu_{d} = F/W$ in stationary sliding at
$V_{0} = 10 \, \mu$m/s, at successive times $t_{i}$ ranging from $\sim 10$ to
$\sim 10^{5}$ seconds. In between two measurements, the driving is
stopped. After such a stop, the slider slows down, and
ultimately stops (after a few seconds; see Figure \ref{fig:3}) at a finite
force level $F_{stop}$ where no creep motion is detectable any
longer. As shown on Figure \ref{fig:2}, $\mu_{d}(V_{0})$ is found to increase
quasi-logarithmically with the time $t$ elapsed from the moment of
first contact. We have checked that $\mu_{d}(V_{0}\mid t)$ is
independent of the number and distribution of the $t_{i}$'s.

This result can be attributed to the geometric aging of a population
of contacts {\it which retain their identity during sliding}, in
agreement with observations from optical imaging \cite{DK} of the
sliding interface \cite {Kpc}. Indeed, in this case, the geometric age
of the microcontacts simply reads $\Phi = t$, and :
\begin{equation}
\label{area}
\Sigma_{r}(t) = \Sigma_{0} [1+ m \ln(1 + \frac{t}{\tau})]
\end{equation}
For PMMA at room temperature, $\tau$ has been estimated \cite{PRB1} to be at most
of order $10^{-2}$ s, and $m \approx 10^{-2}$.

The measured logarithmic slope, $d\mu_{d}(V_{0}\mid t)/d(\ln t) =
2.3\, 10^{-2}$, yields a time evolution of $\Sigma_{r}$ compatible with that
deduced from the increase with waiting time of the static friction
coefficient of a \{rough/rough\} PMMA interface: 
$d\mu_{s}/d(\ln t) = 1.7\times 10^{-2} \pm 4\times 10{-3}$.

The situation encountered here is to be contrasted with the
rough/rough configuration, in which sliding permanently destroys the
microcontacts and replaces them by fresh ones. The use of a rough/flat
system enables us  to take advantage of the quasi-saturation
associated with logarithmic growth, by performing experimental runs of
limited duration $\Delta t$ on an interface of geometric age $\Phi >>
\Delta t$. Typically, for all the results presented below, $\Phi$
ranges from $\sim 10^{4}$ to $\sim 10^{5}$ s, while $\Delta t < 1$
hour. $\Sigma_{r}$ variations are then negligible, giving us access to
the fine variations of the ``rheological stress" $\sigma_{s}$.

\section{Experimental results}
\label{sec:2}

\subsection{Friction force level versus silanization}
\label{subsec:2.1}
A first, rough characterization of the effect of the physico-chemical
state of the glass surface is provided by the value $\bar\mu$ of $\mu _{d}$
measured at some fixed $V_{0}$. Indeed, the friction force
variations to be described below, which provide insight into the
frictional joint dynamics, are always quite small as compared with
the average level.

First of all, as can be expected for such a highly wettable surface,
bare glass tracks yield highly scattered $\bar\mu$-levels, ranging
between $0.4$ and $0.7$

Silanization results in a sizeable decrease of both the average
fric\-tion level and its scat\-ter. Namely, for D1 tracks, $\bar\mu$ lies
between $0.25$ and $0.36$. For more thoroughly sil\-anized, D5
samples, $\bar\mu$ ranges from $0.13$ to $0.2$ -- a very low level for
unlubricated friction \cite{BT}.

It is worth mentioning that contact angle measurements performed with
pure water drops hardly permit to distinguish between D1 and D5
surfaces. The contact angle for both types of tracks is $85^{\circ} \pm
5^{\circ}$, compatible with previous measurements for short chain
silanes \cite{JMB}. This points towards the high
sensitivity of solid friction as a probe of small scale chemical
inhomogeneity of extended surfaces.

In view of the poor reproducibility of the results on nominally bare
surfaces, clearly resulting from the presence of uncontrolled
adsorbates, most of the results reported below have been obtained with
D1 and D5 silanized glass tracks.

\subsection{Static friction peak as a proof of structural aging}
\label{subsec:2.2}
\begin{figure}[htbp]
    $$
    \includegraphics[width=70mm]{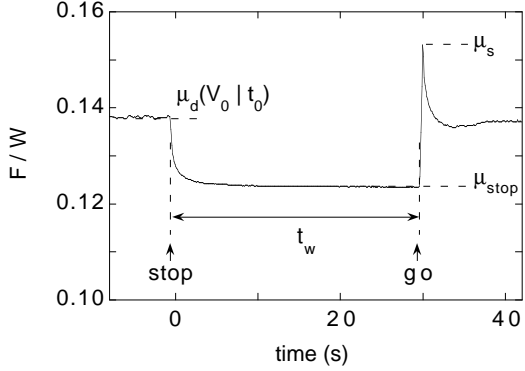}
    $$
    \caption{Force trace corresponding to a stop-and-go experiment 
    (see text), on a D5 track. $V_{0}=10\, \mu$m.s$^{-1}$, 
    $t_{0}=10^{4}$ s. $\mu_{stop}$ corresponds to the jamming stress 
    level; $\mu_{s}$ measures the static threshold.}
    \label{fig:3}
\end{figure}

We prepare an ``old" interface, with geometric age $t_{0}$, on which
we then perform a set of ``stop-and-go" experiments. That is, starting
from steady sliding at velocity $V_{0}$, the drive is stopped, and
the slider left at rest under its self-selected tangential force
$F_{stop}$ (see \ref{fig:3}).
After waiting a time $t_{w}$ after the drive has been stopped, loading
is resumed at the same $V_{0}$.
The corresponding force record, shown on Figure \ref{fig:3}, exhibits a
so-called static friction peak. Namely :

(i) In a first regime, $F$ increases linearly, at the rate $KV_{0}$
corresponding to the elastic loading of the non sliding system. The
interface is pinned.

(ii) This static regime ends at the peak value $F_{s} = \mu_{s}W$,
which defines the static friction coefficient $\mu_{s}$.

Note that, strictly speaking, due to the finite loading stiffness $K$,
sliding starts (depinning occurs) at a slightly lower force level.
Indeed, when the peak is reached, necessarily, the instantaneous
sliding velocity $\dot x$ equals the driving one since, inertia being
negligible, $K(V_{0} - \dot x) = dF/dt = 0$ at this point. Close
inspection of the $F$-trace indeed reveals, in the immediate vicinity
of $F_{s}$, a small
rounding off.

(iii) Beyond the static threshold, a rapid force drop is~observed, and $F$
reaches the steady state plateau value $\mu_{d}(V_{0}\vert t_{0})W$.

\begin{figure}[htbp]
    $$
    \includegraphics[width=70mm]{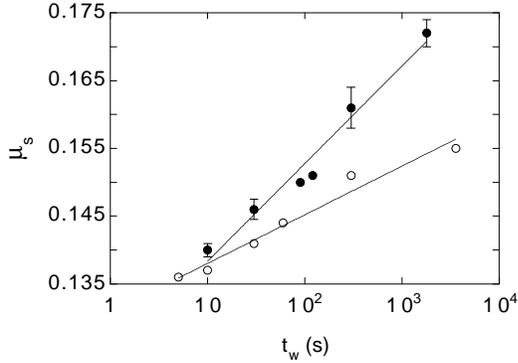}
    $$
    \caption{Growth of the static friction coefficient with waiting 
    time $t_{w}$, on a D5 track. ($\circ$): waiting under zero shear stress 
    ($\mu_{w}=0$). ($\bullet$): waiting at the jamming level 
    $\mu_{stop}$.}
    \label{fig:4}
\end{figure}

Moreover, when varying the waiting time $t_{w}$ over
more than two decades, we observe (fig.\ref{fig:4}) that $\mu_{s}$
increases quasi-logarithmi\-cally with  $t_{w}$: as
time lapses, the interfacial pinning strength  of the non moving
system increases. Since this occurs at constant $\Sigma_{r}$, we must
conclude that, when at rest, the adhesive joints experience {\it
structural aging}.

A second important conclusion results from the presence of the
post-peak stress drop in the transient characteristics. Indeed, this
decrease of the interfacial strength indicates that, conversely, when
sliding sets in, the ``structural age" $\varphi$ decreases from its initial
value,
reached after waiting $t_{w}$, to the smaller value $\varphi(V_{0})$
, reached when steady motion is established : {\it motion
rejuvenates the adhesive joints}.

Structural aging is a robust phenomenon, which we have observed with
all interface samples, involving either bare or silanized glass tracks.

The aging dynamics can be characterized by the value of the
logarithmic slope $\beta_{s} = d\mu_{s}/ d(\ln t_{w})$. This is found
to lie in the range $1.5\times 10^{-3}$ --$ 5\times 10^{-3}$. Although this level of
dispersion is well above experimental uncertainty, no correlation can
be evidenced between $\beta_{s}$ values and $\bar\mu$ levels.

\subsection{Dependence of aging on shear stress level at rest}
\label{subsec:2.3}
In the stop-and-go experiments described above, the system ages at
rest under the finite shear force $F_{stop}=\mu_{stop}W$ at which
its self decelerating creep becomes immeasurably slow. This ``jamming
force level" is found to lie at most $10\% $ below the steady one
$\mu_{d}W$ (see for example Figure \ref{fig:3}).

The question then naturally arises of whether the shear stress level
at rest affects the aging dynamics. For this purpose we have extended
the standard stop-and-go protocol as follows. At time $t = 0$ the
drive is reversed, so as to unload the system down to the desired
``waiting" level $\mu_{w}W$, with $0 \leq \mu_{w} < \mu_{stop}$. After
waiting $t_{w}$, we measure the height of the static peak
$\mu_{s}(t_{w}\vert \mu_{w})$.

Figure \ref{fig:4} shows the variations with waiting time of
$\mu_{s}(t_{w}\vert \mu_{stop})$ and $\mu_{s}(t_{w}\vert 0)$ for a D5
sample. The corresponding logarithmic slopes are respectively
$\beta_{s}(\mu_{stop}) = 6.2\times 10^{-3}$ and $\beta_{s}(0) =
3.7\times 10^{-3}$. Clearly, the aging dynamics is markedly accelerated by
the presence of the shear stress. This effect is observed with all
samples.

\begin{figure}[htbp]
    $$
    \includegraphics[width=70mm]{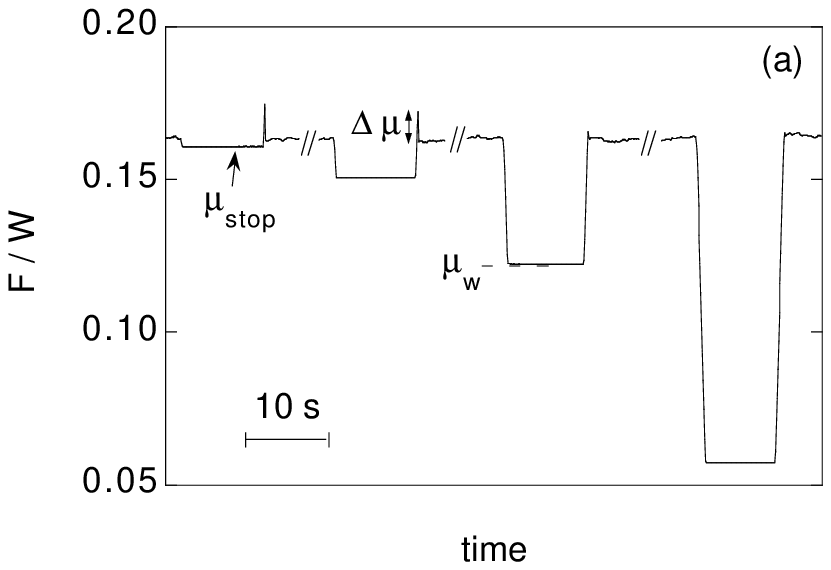}
    $$
    $$
    \includegraphics[width=72mm]{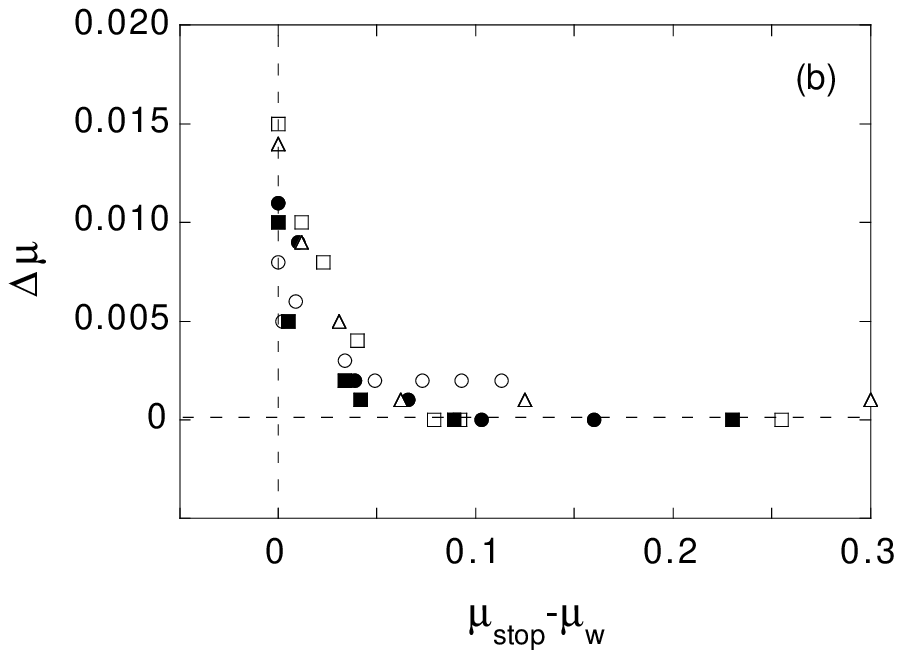}
    $$
    \caption{(a) Force trace for a set of four stop-and-go runs on a 
    D5 track at 
    $V_{0}=10\, \mu$m.s$^{-1}$. Successive stops all last for 
    $t_{w}=10$ s, and are performed under decreasing $\mu_{w}$ levels.
    (b) Post static peak reduced stress drop $\Delta\mu$ versus 
    $\mu_{stop}-\mu_{w}$, for three D1 ({\tiny $\blacksquare$}, {\tiny $\square$}, 
    {\tiny $\bigtriangleup$}) and two D5 ($\bullet$, $\circ$) tracks.}
    \label{fig:5}
\end{figure}

It can be further qualified by measuring $\mu_{s}$, for a
given waiting time $t_{w}$, at various $\mu_{w}$ levels. The result
of such a set of experiments, performed with four different silanized
surfaces, is shown on Figure \ref{fig:5} for $t_{w} = 10$ s. In view of
the spread of average friction levels, we have plotted, for the sake of
comparison, the drop $\Delta\mu$ associated with the post peak stress
drop, versus $(\mu_{stop} - \mu_{w})$. It is seen that:
\begin{itemize}
    \item To a good approximation, all data collapse on a single curve --
    indicating, again, a very weak correlation between aging dynamics
    and average friction level.
    \item The structural age reached at $t_{w}$ decreases very rapidly as
    the waiting stress level decreases below the jamming one.
\end{itemize}

\begin{figure}[htbp]
    $$
    \includegraphics[width=70mm]{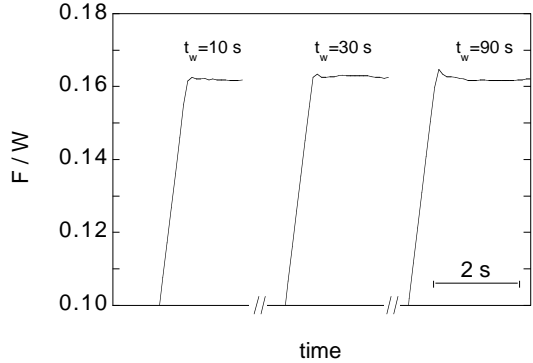}
    $$
    \caption{Three force records for a D5 track, illustrating the 
    behaviour when reloading after stops at $\mu_{w}=0$ of increasing 
    duration $t_{w}$. The static peak becomes visible after a latency 
    time which, here, lies between 30 s and 90 s.}
    \label{fig:6}
\end{figure}

Note that some data correspond to $\Delta\mu = 0$, meaning that, in
such cases, no static peak was observable after waiting $t_{w} = 10$ s. In
these cases, upon loading, the end of the static regime is revealed
by a break in the slope of the $F(t)$ trace. For these samples at such
low $\mu_{w}$, the static peak starts developing only for larger
$t_{w}$ (see Figure \ref{fig:6}). Systematic investigations at 
$\mu_{w} = 0$ reveal that the
corresponding ``latency time" is strongly sample-dependent. We have
observed values ranging from less than $10$ s to more than $2000$ s.

\subsection{Velocity dependence of the steady sliding friction
coefficient}
\label{subsec:2.4}

\begin{figure}[htbp]
    $$
    \includegraphics[width=70mm]{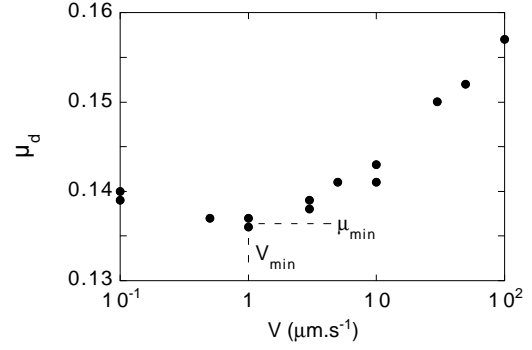}
    $$
    \caption{Dynamic friction coefficient versus velocity $V$ for a 
    D5 sample.}
    \label{fig:7}
\end{figure}

We have also systematically measured the $\mu_{d}(V)$ characteristics
of our geometrically old interfaces. A typical result is shown on
Figure \ref {fig:7}. $\mu_{d}(V)$ exhibits a minimum $\mu_{min}$ at $V =
V_{min}$. For this D5 sample, $V_{min} \approx 1 \mu$m/s.

The velocity weakening observed for $V < V_{min}$ provides another evidence
that the structural age $\varphi$ decreases with $V$, i.e. that the
system rejuvenates when sliding.

For $V > V_{min}$, $\mu_{d}$ exhibits clearly a logarithmic increase,
reminiscent of that found for rough/rough (e.g. PMMA) interfaces. It
has been shown that this effect can be attributed to thermally
activated premature depinning of bistable centers with a volume on
the order of a cubic nanometer. These centers would thus be the basic
dissipative units within the sheared adhesive joint. It is this
mechanism, which should also be at work here, which yields the
$V$-dependence of the sliding stress given by equation (\ref{eq:rheo}).
For PMMA at $300$K, the dimensionless parameter  appearing in
this equation $\alpha \approx 5\times 10^{-2}$.

\begin{figure}[htbp]
    $$
    \includegraphics[width=70mm]{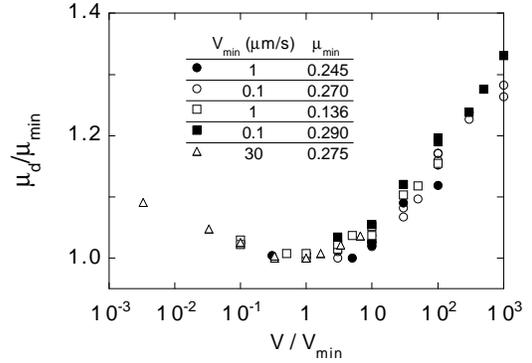}
    $$
    \caption{Reduced (see text) dynamic friction coefficient 
    $\mu_{d}/\mu_{min}$ versus reduced velocity $V/V_{min}$ for D1 
    ($\bullet$, $\circ$, {\tiny $\blacksquare$}, {\tiny 
    $\bigtriangleup$}) and D5 ({\tiny $\square$}) tracks.}
    \label{fig:8}
\end{figure}

For the sake of comparison, we have rescaled our $\mu_{d}(V)$ data
by $\mu_{min}$. The result, for five different (D1 and D5) samples is
plotted on Figure \ref{fig:8}, in which $V$ has also been scaled by
$V_{min}$. It is seen that, for $V/V_{min}>1$, all data then collapse on a single curve. The common
reduced logarithmic slope in the velocity-strengthening regime is
$5\times 10^{-2} \pm 6\times 10^{-3}$, in full agreement with the previously
evaluated $\alpha$ for PMMA/PMMA.

 From this, one may also conclude that, already rather close above
$V_{min}$, the sliding induced rejuvenation effect is
quasi-saturated (which justifies a posteriori our choice of
$\mu_{min}$ for the rescaling).

Note that, while the $\mu$-levels for the various samples differ at
most by a factor of $2$, $V_{min}$ values span more than two decades
(from $0.1$ to $ 30 \mu$m/s). This considerable scatter must certainly
be related to the fact that aging/rejuvenation is associated with
activated ``jumps" above the barriers of a glass-like energy landscape.
Hence, roughly speaking, an exponential sensitivity to variations of
this landscape.

Velocity-weakening is well known to promote stick-slip instabilities.
Indeed, we did observe stick-slip oscillations for silanized samples,
but only for of few of the D1 ones, and in the lower part of their
$V$-weakening regime. Let us recall that, as long as the rejuvenation
mechanism responsible for weakening is non-instantaneous, a negative
$d\mu_{d}/dV$ does not necessarily entail stick-slip, the presence of
which is governed by the detailed dynamics of the age variable and by
the external control parameters, e.g. the external loading stiffness.

It is worth mentioning that, with bare glass tracks, we always observe
stick-slip, up to a strongly sample-dependent velocity threshold,
ranging from $\sim 10 \mu$m/s to $> 100 \mu$m/s. That is, as a rule
of thumb, the higher the average friction level, the stronger the
trend towards destabilization of steady sliding. However, the poor
reproducibility of systems involving bare glass does not permit to
make this statement more quantitative.

\subsection{Long lasting transients}
\label{subsec:2.5}

\begin{figure}[htbp]
    $$
    \includegraphics[width=70mm]{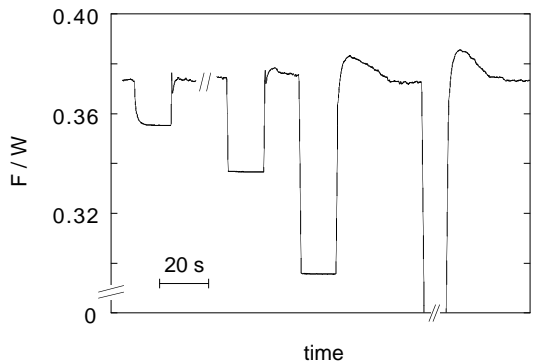}
    $$
    \caption{Long lasting bump shaped transients, after four stops of 
    duration $t_{w}=20$ s, illustrating the increase of transient bump 
    amplitude with decreasing $\mu_{w}$ level.}
    \label{fig:9}
\end{figure}

While, for D5 tracks, the transient behavior only consists of the
above described static peaks, for D1 systems a more complicated
dynamics emerges. Indeed (see Figure \ref{fig:9}), in most cases, the
force traces also exhibit long lasting post-peak bump shaped
transients. Comparison of data for various driving velocities
spanning about a decade enable us to assert that this slow transient
is not governed by a characteristic time but, rather, by a slid length
lying in the 100 -- 300 $\mu$m range. This must be contrasted with the
distance slid during an initial stress drop, which never
exceeds 1 $\mu$m, indicating that the associated mechanisms are of a
different nature.

Moreover, the bump amplitude is found to increase markedly as the
waiting stress level is decreased (Figure \ref{fig:9}).

These elements lead us to associate long lasting transients with
anelastic evolution of the slider PMMA mol\-ecules forming the joints:
when the stress level is stepped down, as occurs during a stop, they
are likely to experience strain recovery. Upon reloading, this process
is reversed, and the corresponding work must be provided by the
external force.

That this mechanism is attributable to conformational evolution of the
surface region {\it of the slider} is proved by the fact that these
long transients are most pronounced when using freshly lapped PMMA
surfaces. Bump heights then decrease gradually with cumulated slid
distance, on scales of the order of tens of centimeters. Large bumps
reappear after relapping. This suggests that repeated stress cycling
leads to irreversible alignment of the polymer molecules -- an effect
documented already long ago by Tabor {\it et al} \cite{tabor2}, in particular with PTFE,
for which it is most pronounced.

\section{Discussion}
\label{sec:3}

The set of experimental results reported in Section \ref{sec:2} leads
us to assert that the nanometer-thick adhesive joints in which
frictional sliding localizes can be considered in all respects as a
confined 2D soft glass medium, or, equivalently, threshold fluid. Namely :

(i) Below a static threshold, it responds elastically: it is pinned
in a disordered solid state induced by confinement under the normal
stress.

(ii) This static threshold increases logarithmically with the waiting
time : when at rest, the joint ages.

(iii) When sliding starts, the stress drops until reaching its
(lower) stationary level : motion rejuvenates the joint.

(iv) This rejuvenation corresponds to the existence of a
velocity-weakening regime for dynamic friction. As $V$ increases,
rejuvenation saturates and friction becomes velocity-strengthening.

Moreover, we find that the (logarithmic) aging rate increases
markedly with the shear stress applied to the system at rest. This
influence on aging of stress during waiting is akin to the recent
results obtained by Viasnoff {\it et al} \cite{lequeux2} with a colloidal glass. They have
shown that applying to this system, aging under stress-free
conditions, an oscillating shear stress of finite duration does
result in a strong perturbation of the aging process.

In our
stop-and-go experiments, the whole aging process takes place under
constant shear below the static threshold. This situation is reached
after performing a ``mechanical quench", during which sliding
decelerates and stops at the prescribed level. Note that the
duration $\Delta t$
of the quench depends on the final stress level : for the largest
one, corresponding to the jamming level $\mu_{stop}$, it is determined
by the sliding dynamics itself, and is typically a few seconds. When
we unload below $\mu_{stop}$, $\Delta t$ is controlled by the
unloading time -- at most $1$ second. For some samples at $\mu_{w} =
0$, no static peak is observed for a finite latency period. This
indicates that, in these experiments (i) unloading
has been fast enough for the joints to be quenched at the initial
age $\varphi (V_{0})$ they had when sliding; (ii) the subsequent
aging dynamics is slow enough that the peak growth is too small to be
measured up to the latency time.

The aging rate increase with waiting stress level is most likely
ascribable to the associated biasing of the energy landscape and to the
corresponding decrease of ``trap depths" \cite{sollich1,sollich2}. This effect might be
responsible in part for the observation by Berthoud {\it et al} 
\cite{PRB1} that the
logarithmic growth rate of the static peak height of a
\{rough/rough\} PMMA interface is larger under jamming than under
stress-free conditions. It would be of interest that analogous studies
be performed with
3D threshold fluids such as colloidal glasses.

While the above described phenomenology of friction exhibits all the
main qualitative features of soft glass rheology, when inspected more
closely, our results lead to two further questions.

(i) The logarithmic velocity dependence of the dynamic friction
coefficient $\mu_{d}$ in the strengthening regime is fully compatible
with a description in terms of premature depinning events induced by
the thermal noise, as discussed in references \cite{PRB2,persson}.
Existing theories of sheared glasses \cite{sollich1,sollich2,kurchan}
and of plasticity of amorphous
media \cite{argon,langer} lead to the idea that, while thermal noise should be relevant
at low shear rates, beyond some crossover, the dynamical noise
associated with the depinning events themselves should take over. No
prediction is available for the moment about the position of this
crossover.

In our experiments, the $V$-strengthening regime
corresponds to shearing rates $\dot{\gamma} = V/h
\simeq 10^{3}$--$10^{5}$. This seems to indicate that, for frictional
joints, the $\dot{\gamma}$ crossover value  would be at least that
huge. Whether this rather surprising result is of some degree
of generality, or is due
to possible screening of elastic interactions related with the
confinement of our 2D medium remains an open question.

(ii) Average friction levels are highly sensitive to the chemical
differences between surfaces obtained under nominally identical
silanization conditions. This is most probably related with the fact
that defects in the silane coverage, however small their scale,
leave non passivated high energy, strong pinning sites on the glass
surface. These lead to strong corrugations of the confining potential,
hence of the glass-like energy landscape. The impossibility to
correlate our $\beta_{s}$ and $\bar{\mu}$ data leads us to suspect that,
while $\bar{\mu}$ measures an average pinning strength, $\beta_{s}$
might be sensitive to much finer details of the landscape. Numerical
modelling would probably be very helpful to shed light on this question.

\begin{acknowledgement}
    We are indebted to J.M. Berquier and P. Silberzan for advice 
    about glass silanization. We thank L. Leger, L. Cugliandolo and 
    J. Kurchan for illuminating discussions.
    
\end{acknowledgement}

%\bibliographystyle{unsrt}
%\bibliography{epje}
%

\end{document}